\documentclass[twocolumn,showpacs,preprintnumbers,amsmath,amssymb]{revtex4}

\usepackage{graphicx}
\usepackage{dcolumn}
\usepackage{bm}

\begin{document}

\title{Renormalization of spin-orbit coupling in quantum dots \\
	due to Zeeman interaction}

\author{Manuel Val\'{\i}n-Rodr\'{\i}guez}
\affiliation{%
Departament de F\'{\i}sica, Universitat de les Illes Balears,
E-07122 Palma de Mallorca, Spain
}%

\date{March 26, 2004}

\begin{abstract}
We derive analitycally a partial diagonalization of the Hamiltonian representing
a quantum dot including spin-orbit interaction and Zeeman energy on an equal
footing. It is shown that the interplay between these two terms results in a
renormalization of the spin-orbit intensity. The relation between this feature
and experimental observations on conductance fluctuations is discussed, finding
a good agreement between the model predictions and the experimental behavior.
\end{abstract}


\maketitle

In recent years, the physics concercing the spin of electrons has become a
widely studied subject in the context of semiconductor physics. The fundamental
physics involved and the possibility of future technological applications
motivate this research \cite{wolf,datta,loss,wang,schlie,pershin}.

In particular, the role of the spin in low-dimensional semiconductor structures,
like quantum dots, is especially attractive because of the ability to
control the relevant parameters of the nanostructures available at present. Thus,
allowing for the knowledge of the interactions governing the spin dynamics
within these structures.

Among the interactions concerning the spin, a relevant intrinsic interaction
in non-magnetic semiconductors is the spin-orbit coupling, which has been extensively
studied in the context of quantum dots \cite{naz,des,cheng,vos,gov}. This interaction stems from
the relativistic correction to the electronic motion. The electric fields present in
the semiconductor are felt by the electrons in their intrinsic reference frame as a
spin-dependent magnetic field, whose intensity depends on the momentum of the particle.
Depending on the origin of the electric field there are distinguished two sources of
spin-orbit coupling: the lack of inversion symmetry in a bulk semiconductor originates
the so-called Dresselhaus term \cite{dress}, whereas the asymmetry of the potential
confining the two-dimensional electron gas of a heterostructure is the responsible of
the Bychkov-Rashba interaction \cite{rash}.

The other spin-dependent interaction of relevance for the purposes of this letter is
the Zeeman interaction, arising from the direct coupling with an applied magnetic field 
of the intrinsic magnetic moment associated with the spin.

It is the aim of this letter to treat, in an analytical way, the correlation
between Zeeman and spin-orbit interactions in a quantum dot, which leads to an amplification 
of the spin-orbit effects.

Analytical expressions for the level structure of quantum dots with spin-orbit couping, obtained through 
perturbation theory, already exist in the literature \cite{gov,vala}. However, these calculations rely
on a regime differentiation where the energy scale associated with Zeeman interaction
is much bigger than that corresponding to the spin-orbit coupling or vice versa. In
a recent letter, Aleiner and Fal'ko \cite{alei} introduced a method that takes advantadge of the
relative weakness of the spin-orbit interaction in order to treat analitically
this interaction in quantum dots. The method consists in the application
of a proper unitary transformation to the Hamiltonian that diagonalizes its spin-dependence
to second order in the spin-orbit intensities (higher order corrections are negligible
because of the implicit weakness of the spin-orbit interaction). It is shown that, in the 
transformed reference frame, the spin-orbit interaction can be interpreted to produce
an spin-dependent Aharanov-Bohm flux. Nevertheless, this procedure involves, exclusively,
the diagonalization of the spin-orbit terms, without taking into account the 
correlation of these terms with the Zeeman interaction.

In order to characterize the system under study, we consider the motion of the electrons
to be restricted to the two-dimensional confining region of a quantum well. In contrast
to the approach of Aleiner and Fal'ko, which was valid for any lateral confining potential,
it is needed to specify the coordinate-dependence of the in-plane potential in order to
treat the spin-orbit coupling and the Zeeman interaction simultaneously. For this purpose, 
we shall use a harmonic oscillator potential.
\begin{equation}
V(x,y)=\frac{1}{2}m^*\omega_{0}^2(x^2+y^2)
\end{equation}

where $m^*$ represents the electronic effective mass in the conduction band and $\omega_0$
is the natural frequency of the potential, that provides a measure of the energy scale and
the size of the system.

We also include the effect of a magnetic field oriented perpendicular to the plane of the
quantum well, consequently, orbital in-plane motion will be affected by this applied field.
The orbital role of the magnetic field is taken into account through the vector potential-
dependence of the generalized momentum $\vec{P}=\vec{p}+\frac{e}{c}\vec{A}$ where 
$\vec{A}=\frac{B}{2}(-y,x,0)$ is the vector potential in the symmetric gauge. In this way,
the spin-independent part of the Hamiltonian reads,
\begin{equation}
{\cal H}_0=\frac{1}{2m^*}(P_x^2+P_y^2)+\frac{1}{2}m^*\omega_0^2(x^2+y^2)
\end{equation}

Zeeman interaction corresponding to the vertical magnetic field and spin-orbit coupling
constitute the spin-dependent part of the Hamiltonian. For simplicity, we shall consider
here only the Bychkov-Rashba term,
\begin{eqnarray}
{\cal H}_z=\frac{1}{2}g^*\mu_BB\sigma_z=\frac{\varepsilon_z}{2}\sigma_z\\
{\cal H}_R=\frac{\lambda_R}{\hbar}\left(P_y\sigma_x-P_x\sigma_y\right)
\end{eqnarray}
\begin{equation}
{\cal H}={\cal H}_0+{\cal H}_R+{\cal H}_z
\end{equation}

where $g^*$ represents the effective gyromagnetic factor, $\mu_B$ is the Bohr's magneton,
$\lambda_R$ the zero-field Bychkov-Rashba intensity and the different $\sigma$'s represent
the Pauli matrices.

To achieve the diagonalization of the above Hamiltonian we use an approach similar to that
introduced by Aleiner and Fal'ko \cite{alei}. By means of an unitary transformation, the Hamiltonian is 
diagonalized in spin space up to second order in the spin-orbit and Zeeman parameters. It
is important to mention that the energy scale corresponding to the spin-orbit and Zeeman 
interactions must be much smaller than that corresponding to the spin-independent
Hamiltonian (${\cal H}_0 \gg {\cal H}_R,{\cal H}_z$), in order to neglect high order terms
introduced by the unitary transformation. Nevertheless, no restriction is imposed to the 
relative importance of the two spin-dependent interactions, i.e., under the assumption of
spin-dependent effects to be much smaller than spin-independent ones, the relative weight
between Zeeman and spin-orbit interactions becomes irrelevant in what concerns to the
suitability of the method.

The Hamiltonian in the transformed reference frame is given by
\begin{equation}
\label{eqone}
{\cal H}^{\prime}={\cal U}^{\dagger}{\cal H}{\cal U},\:\:{\cal U}=e^{-i\beta\hat{O}}
\end{equation}
The explicit dependence on coordinates and momenta of the unitary transformation reads,
\begin{equation}
\label{eqtwo}
\hat{O}=\left(P_x-\frac{\hbar m^*\omega_0^2}{\varepsilon_z}\,y\right)\sigma_x+ 
		\left(P_y+\frac{\hbar m^*\omega_0^2}{\varepsilon_z}\,x\right)\sigma_y 
\end{equation}
and,
\begin{equation}
\label{eqthr}
\beta=-\frac{\lambda_R}{\hbar}\frac{\varepsilon_z}{(\hbar\omega_0)^2+(\hbar\omega_c)\varepsilon_z - \varepsilon_z^2}
\end{equation}

All the parameters in these formulas have been previously introduced except the cyclotron
frequency $\omega_c=eB/m^*c$ which quantifies the effect of the vertical magnetic field in
the orbital motion of the electrons.

We expand the transformed Hamiltonian given by Eq. \ref{eqone} up to second order 
in '$\beta$' using the Baker-Haussdorf lemma,
\begin{equation}
{\cal H}^\prime={\cal H} + i\beta\left[\,\hat{O},{\cal H}\,\right]-\frac{\beta^2}{2}\left[\,\hat{O},
		\left[\,\hat{O},{\cal H}\,\right]\right]+ o(\beta^3)
\end{equation}

The resulting Hamiltonian is diagonal in spin space if the smaller terms $o(\beta^3)$ are neglected,
\begin{eqnarray}
\label{htrans}
{\cal H}^\prime \approx \frac{1}{2m^*}
\left(1-\frac{2\lambda_R^2\frac{m^*}{\hbar^2}\varepsilon_z}{(\hbar\omega_0)^2+(\hbar\omega_c)\varepsilon_z-\varepsilon_z^2}\,\sigma_z\right)(P_x^2+P_y^2)+\nonumber \\
+\frac{1}{2}m^*\omega_0^2(x^2+y^2)-
\lambda_R^2\frac{m^*}{\hbar^2}
\left(1-\frac{\varepsilon_z^2}{(\hbar\omega_0)^2+(\hbar\omega_c)\varepsilon_z}\right)^{-1 }+ \nonumber \\
+\frac{\varepsilon_z}{2}\sigma_z
-\lambda_R^2\frac{m^*}{\hbar^3}\left(1+\frac{\varepsilon_z(\hbar\omega_c-\varepsilon_z)}{(\hbar\omega_0)^2}\right)^{-1}
(xP_y-yP_x)\sigma_z \nonumber \\
\end{eqnarray}

Note that, the above transformed Hamiltonian can be straightforwardly solved for
the two spin eigenstates (in the intrinsic frame) by comparison with the well-known 
two-dimensional harmonic oscillator problem, responsible of the Fock-Darwin levels.

A first glance at the result of Eq. \ref{htrans} shows that, in the transformed
frame, the full Hamiltonian is diagonal in the spin basis where the Zeeman term was already 
diagonal, the $\sigma_z$-basis. Then, it could result quite surprising to include the Zeeman
parameter in the unitary transformation if the Zeeman interaction was already diagonal in
the final spin basis. However, note that the present approach contains
rather subtle aspects: the unitary transformation given by Eqs.\:\ref{eqone},\ref{eqtwo},
\ref{eqthr}, apart of the spin-orbit term , also diagonalizes
crossed terms that depend both on spin-orbit and Zeeman parameters; thus, containing the interplay
between these two interactions (Zeeman term is already diagonal).

Let us take a especial care in the understanding of the term
\begin{equation}
-\lambda_R^2\frac{m^*}{\hbar^3}\frac{1}{1+\frac{\varepsilon_z(\hbar\omega_c-\varepsilon_z)}
{(\hbar\omega_0)^2}}(xP_y-yP_x)\sigma_z 
\end{equation}

that appears in the transformed Hamiltonian. Following the approach of Aleiner and
Fal'ko for a parabolic dot \cite{lolo2} the correction introduced by spin-orbit interaction is given
by
\begin{equation}
-\lambda_R^2\frac{m^*}{\hbar^3}(xP_y-yP_x)\sigma_z 
\end{equation}

Note that, both terms are very similar and the only difference between them is the value of
their intensity. This means that, when the interplay between Zeeman energy and spin-orbit
coupling is considered, the effective intensity of the spin-orbit interaction actually depends
on the magnetic field.
\begin{equation}
\label{nucleo}
\lambda_{eff}^2=\frac{\lambda_R^2}{1+\frac{\varepsilon_z(\hbar\omega_c-\varepsilon_z)}{(\hbar\omega_0)^2}}
\end{equation}

Both orbital and magnetic terms, represented by the cyclotron frequency and the Zeeman
parameter respectively, grow linearly with the magnitude of the applied field. The relative 
smallness of the effective gyromagnetic factor in certain materials, such as GaAs or AlGaAs,
supposes that the energy associated to the orbital magnetic term is always much bigger
than the corresponding to the Zeeman energy ($ \varepsilon_z \ll \hbar\omega_c $).
Therefore, Eq. \ref{nucleo} can be approximated without significant error by:
\begin{equation}
\label{nucleo2}
\lambda_{eff}^2=\frac{\lambda_R^2}{1+\frac{\hbar\omega_c\varepsilon_z}{(\hbar\omega_0)^2}}
\end{equation}

The product between Zeeman  parameter and the cyclotron frequency appears as the main
parameter determining the effective intensity of the spin-orbit coupling in the dot. It is important
to note that the sign of the gyromagnetic factor determines if the effective spin-orbit intensity is
enhanced or reduced as the applied magnetic field is increased, thus, providing a quantity
whose effects can be measured and that reflects unambiguously the sign of '$g^*$'.

Despite of the cyclotron frequency appears in the effective intensity given by Eq. \ref{nucleo}, its 
origin is a consequence the Zeeman interaction. If Zeeman energy is set 
to zero, the orbital magnetic terms do not modify the spin-orbit intensity. 
However, if we supress the orbital effects of the magnetic field by setting 
the cyclotron frequency to zero; thus, only conserving the Zeeman term, there
is still an effective modification of the spin-orbit intensity. 
\begin{equation}
\label{eqlambda}
\lambda_{eff}^\prime=\frac{\lambda_R}{\sqrt{1-(\frac{\varepsilon_z}{\hbar\omega_0})^2}}
\end{equation}

This result can be interpreted as a renormalization of the spin-orbit intensity
induced by its correlation with the Zeeman interaction.

It is important to note that, in this case, the B-evolution of the effective intensity is
much less pronounced than that given by Eq. \ref{nucleo2} due to the bigger
B-dependence of the cyclotron frequency stated before. Note also that the
confining potential plays a relevant role in the amplification of the effective spin-orbit
intensity, since the relative weight between Zeeman energy and the confinement's
energy quantum determines its variation respect to the zero-field value '$\lambda_R$'.
In the limit of strong confinement ($\varepsilon_z \ll \hbar\omega_0$) the effect
vanishes, whereas in the opposite limit of weak confinement, the enhancement of
the effective intensity becomes bigger as the confining potential is softened.
Simultaneously,  as the ratio between '$\varepsilon_z$' and '$\hbar\omega_0$' is
increased, the approach loses its applicability, since the main hypothesis formulated
was the orbital energy scale to be much bigger than that corresponding to the spin-dependent
interactions. The only weak confined systems ($\hbar\omega_0\le \varepsilon_z$) 
which can be treated within this model are those where a significant orbital magnetic
effect is present ($\hbar\omega_c\gg \varepsilon_z$). In this case, the bulk limit 
corresponding to the Landau levels in the presence of spin-orbit coupling can even
be explored, finding a good approximate expression for a problem that actually can
be solved exactly \cite{Schlie}.

The amplification of the spin-orbit intensity due to the Zeeman effect given by
Eq. \ref{eqlambda} is relevant in the determination of the properties corresponding 
to the level structure of the dots. In this sense, conductance fluctuations present in
the transport through the dots is a very valuable tool in order to extract
information about their spectrum \cite{folk,halp,held,falko,crem,brou}.

Recently, Folk {\em et al.} \cite{folk} reported experimental data on the conductance
fluctuations corresponding to open lateral quantum dots created in a GaAs/AlGaAs
heterostructure. They found that the conductance fluctuations were suppressed
as the magnitude of an applied horizontal magnetic field was increased. These
results could be explained if spin-orbit coupling was enhanced by the application
of the horizontal field \cite{halp,folk}. 

The result given in Eq. \ref{eqlambda} is in agreement with these experimental 
observations, since an increase in the Zeeman energy enhances also the effective
value of the spin-orbit intensity. Furthermore, despite of the parabolic potential
is a crude approximation for the confining potential of the dots used by Folk 
{\em et al.}, Eq. \ref{eqlambda} qualitatively describes the experimental dependence
on the size of the system. It was reported that the biggest dots showed a larger suppression
of the conductance fluctuations, up to a factor of suppression $\sim 4$, 
and a factor $\sim 2$ for the smallest ones. The value of this factor
of suppression depends on the degree of enhancement of the spin-orbit coupling with 
the applied field \cite{halp}. Note, from Eq. \ref{eqlambda}, that the
effective spin-orbit intensity also depends on the size of the system through the parameter
$\omega_0$. Bigger dots, characterized by a smaller confining frequencies ($\omega_0$),
would show a bigger enhancement of spin-orbit coupling, for a given Zeeman energy, than
smaller dots, characterized by a bigger frequency. Therefore, the model states that 
larger dots are susceptible to suffer a higher suppression of conductance fluctuations
in agreement with the experimental behavior.

As cited before, the validity of the model is conditioned to the relative
weakness of spin-dependent interactions respect to the potential and kinetic terms.
Nevertheless, taking advantadge of the explicit expression corresponding to the unitary transformation 
a more precise criterion of validity can be established for the model. Note that,
the truncated expansion of the transformed Hamiltonian is consistent when the
operator responsible of the transformation satisfies the condition,
\begin{equation}
\beta\hat{O} \ll 1
\end{equation}
In practice, the model is valid up to the range delimited by $\beta\hat{O}\sim 0.15$ .
The above inequality, in fact, implies two conditions to be simultaneously fulfilled,
\begin{eqnarray}
\left| \frac{\lambda_R\varepsilon_z}{\hbar}\frac{1}{(\hbar\omega_0)^2+(\hbar\omega_c)\varepsilon_z - \varepsilon_z^2}\, P_i \right| \ll 1 \nonumber \\
\left| \frac{\lambda_R m^*}{\hbar^2}\frac{(\hbar\omega_0)^2}{(\hbar\omega_0)^2+(\hbar\omega_c)\varepsilon_z - \varepsilon_z^2}\, x_i \right| \ll 1 
\end{eqnarray}
where '$ P_i $' and '$ x_i $' represent characteristic values
of the electronic momentum and the dimensions of the dot, respectively. The second
condition represents the requirement of weakness for the zero-field spin-orbit 
intensity analogously to ref. \onlinecite{alei}. Meanwhile, the first condition concerns
the range of validity for the Zeeman energy. Note that, there appears the product 
between the Zeeman parameter and the zero-field spin-orbit intensity; therefore, depending
on the smallness of the zero-field spin-orbit intensity, the range of validity for the Zeeman
parameter could become shorter or larger.

In summary, we have accounted analytically for the interplay between spin-orbit coupling 
and Zeeman interaction in a quantum dot model. We have shown that this interplay induces
an effective renormalization of the spin-orbit intensity leading to an amplification of
the spin-orbit effects as the Zeeman energy is increased. Criterions for the validity of 
the model have been established. Comparing these theoretical results with experimental
observations on lateral GaAs/AlGaAs dots, we find that the model contains some features
reveled by the experiments.

This work was supported by Grant No.\ BFM2002-03241 
from DGI (Spain).

\end{document}